\documentclass[a4paper,11pt]{article}
\pdfoutput=1
\usepackage{jheppub} % for details on the use of the package, please
\usepackage{graphicx,color,rotating}
\usepackage{hyperref}
\usepackage{epsfig,color}
\usepackage{slashed}
\usepackage{amsfonts}
\usepackage{ulem}

\newcommand{\gsim}{\raisebox{-0.13cm}{~\shortstack{$>$ \\[-0.07cm] $\sim$}}~}

\usepackage{tabularx}
\usepackage{graphicx}
\usepackage{adjustbox}
\usepackage{tabularx}

\begin{document}
\title{Correlated signals of first-order phase transitions and 
primordial black hole evaporation}

\renewcommand{\thefootnote}{\arabic{footnote}}

\author{
Danny Marfatia$^{1}$ and
Po-Yan Tseng$^{2}$}
\affiliation{
%$^1$ Department of Physics, University of Illinois at Chicago,
%Illinois 60607 USA \\
%%
%$^2$ Physics Division, National Center for Theoretical Sciences,
%Hsinchu, Taiwan \\
%%
$^1$ Department of Physics \& Astronomy, University of Hawaii at Manoa,
2505 Correa Rd., Honolulu, HI 96822, USA \\
$^2$ Department of Physics, National Tsing Hua University,
101 Kuang-Fu Rd., Hsinchu 300044, Taiwan R.O.C. \\
}
%\pacs{14.80.Bn.,14.80.Da,14.80.Ec}
\date{\today}

\abstract{
Fermi balls produced in a cosmological first-order phase transition may collapse to primordial black holes (PBHs) if the fermion dark matter particles that comprise them interact via a sufficiently strong
Yukawa force. We show that phase transitions described by a quartic thermal effective potential with vacuum energy, \mbox{$0.1\lesssim B^{1/4}/{\rm MeV} \lesssim 10^3$}, generate PBHs of mass, $10^{-20}\lesssim M_{\rm PBH}/M_\odot \lesssim 10^{-16}$, and
gravitational waves 
from the phase transition (at THEIA/$\mu$Ares) can be correlated with an isotropic extragalactic X-ray/$\gamma$-ray background 
from PBH evaporation (at AMEGO-X/e-ASTROGAM).
}

\maketitle

\section{Introduction}

Primodial black holes (PBHs), hypothesized to have formed in the early Universe,
are a good dark matter (DM) candidate~\cite{Hawking:1971ei,Chapline:1975ojl,Khlopov:2008qy,Carr:2016drx,Carr:2020gox,Carr:2020xqk,Green:2020jor}.
They offer a compelling explanation~\cite{Clesse:2016vqa,Bird:2016dcv,Sasaki:2016jop} for the gravitational wave (GW) signals observed 
by LIGO/Virgo~\cite{LIGOScientific:2016aoc,LIGOScientific:2016sjg,LIGOScientific:2017bnn}.
PBHs are typically thought to have been produced in
the collapse of overdense regions developed from primordial inhomogeneities seeded by inflation~\cite{Carr:1974nx,Sasaki:2018dmp}. They may also form directly
during a first-order phase transition (FOPT)~\cite{Hawking:1982ga,Moss:1994iq,Konoplich:1999qq,Kodama:1982sf,Gross:2021qgx,Baker:2021nyl}, perhaps through bubble collisions.

In this work, we focus on a novel mechanism in which PBHs are produced in a FOPT
through an intermediate step, wherein macroscopic Fermi balls (FBs) first form during the FOPT and subsequently collapse to PBHs~\cite{Kawana:2021tde}.
FBs originate from the aggregation of fermion DM particles trapped in the false vacuum as the true vacuum bubbles expand~\cite{Hong:2020est,Marfatia:2021twj,Witten:1984rs,Bai:2018dxf}.
The FOPT must take place in a dark sector because in the Standard Model (SM), 
 the electroweak and quantum chromodynamics phase transitions are smooth crossovers. 
We consider the frequently arising quartic effective thermal potential
to generate a FOPT. 
A Yukawa interaction with fermion DM will serve two purposes. The DM particle acquires an additional mass
in the true vacuum, making it heavier than in the false vacuum~\cite{Huang:2022him}. Requiring the DM mass difference to be larger than the critical temperature of the FOPT, forces the DM particles to become trapped in the 
false vacuum. A DM-antiDM asymmetry  then ensures that the shrinking volume of the false vacuum compresses the DM particles into FBs.
Also, the attractive Yukawa force between DM particles causes the FBs to collapse to
PBHs once its range becomes larger than the mean separation distance between particles in the FB~\cite{Kawana:2021tde}. 

%According to this scenario, we found that for $0.1\leq B^{1/4}/{\rm MeV} \leq 10^3$, 
%which describes the energy-potential difference between false and true vacuums at zero temperature, the mass of PBH can be produced within the window 
%$10^{-20}\leq M_{\rm PBH}/M_\odot \leq 10^{-15}$.
Our goal is to show that in a certain mass range of PBHs, gravitational waves from the FOPT and the extragalactic X-ray and GeV $\gamma$-ray background from PBH evaporation can be correlated.
For $M_{\rm PBH}/M_\odot \gsim 10^{-19}$,  PBHs have lifetimes longer than the age of the Universe and are evaporating in the present epoch.
PBH evaporation produces X-rays and GeV $\gamma$-rays that after redshifting are detectable 
at current (SPI, Fermi-LAT) and future (AMEGO-X, e-ASTROGAM) $\gamma$-ray 
telescopes~\cite{e-ASTROGAM:2017pxr,Fleischhack:2021mhc,Laha:2020ivk,Fermi-LAT:2018pfs}.
The corresponding gravitational waves
produced during the FOPT, are detectable
at THEIA~\cite{THEIA}, 
and $\mu$Ares~\cite{Sesana:2019vho}.

The paper is organized as follows. We describe FB formation and the criteria for their collapse to PBHs in section~\ref{sec:FB}. 
We outline our procedure for calculating the PBH evaporation spectra in section~\ref{sec:PBH_evaporation}.
In section~\ref{sec:scan}, we scrutinize the input parameters of the effective potential by performing a parameter scan and calculate
GW and extragalactic photon spectra for a few benchmark points.
We summarize in section~\ref{sec:summary}.

\section{PBH formation}
\label{sec:FB}

We consider the formation of PBHs via a two step process in the hidden sector.
The dark Dirac fermion particles $\chi$'s first aggregate to form macroscopic FBs 
during a dark scalar $\phi$ induced FOPT.
Subsequently, the attractive Yukawa force between $\chi$'s mediated by $\phi$ destabilizes the FBs which collapse to PBHs~\cite{Kawana:2021tde}.

To realize the above scenario, we adopt a model defined by the Lagrangian,
\begin{eqnarray}
\mathcal{L}\supset \bar{\chi}( i \slashed{\partial}-m) \chi  -g_\chi \phi \bar{\chi} \chi -V_{\rm eff}(\phi,T) \,,
\end{eqnarray}
with the finite-temperature quartic effective potential~\cite{Dine:1992wr,Adams:1993zs},
\begin{eqnarray}
V_{\rm eff}(\phi,T)= D(T^2-T^2_0)\phi^2-(AT+C)\phi^3+\frac{\lambda}{4}\phi^4\,,
\end{eqnarray}
which induces a FOPT in the early Universe.
When the temperature drops below the critical temperature $T_c$ 
(defined by $V_{\rm eff}(0,T_c)=V_{\rm eff}(v_\phi(T_c),T_c)$),
the Universe starts to transit from the false vacuum (with $\langle \phi \rangle=0$) 
to the true vacuum (with $\langle \phi \rangle = v_\phi$).
At zero temperature, 
we define $B\equiv |V_{\rm eff}(\tilde{\phi}_+,0)|$ as the vacuum energy density evaluated at the global minimum $\tilde{\phi}_+$~\cite{Marfatia:2021twj}.
Since $V_{\rm eff}(\phi=0,T)=0$, $B$ is the energy density difference between the $\phi=0$ and $\tilde{\phi}_+$ phases at zero temperature.
We choose the input parameters to be
\begin{eqnarray}
\label{eq:input}
\lambda,~A,~B,~C,~D\,,
\end{eqnarray}
and treat the destabilization temperature $T_0$ as a derived quantity~\cite{Marfatia:2021twj}.

An analytical expression for the Euclidean action $S_3(T)/T$ for quartic potentials can be found in Ref.~\cite{Adams:1993zs}. 
We use this to calculate the bubble nucleation rate per unit volume $\Gamma(T)$,
%\begin{eqnarray}
%\label{eq:nucleation_rate}
%\Gamma(T)=T^4 \left( \frac{S_3}{2\pi T} \right)^{3/2} e^{-\frac{S_3}{T}}\,,
%\end{eqnarray}
and hence the fraction of space in the false vacuum $F(t)$~\cite{Marfatia:2021twj}. 
%\begin{eqnarray}
%\label{eq:F_t}
%F(t)={\rm exp}\left[ -\frac{4\pi }{3} v^3_w\int^{t}_{t_c} dt' (t-t')^3 %\Gamma(t') \right]\,,
%\end{eqnarray}
%~\cite{Marfatia:2021twj}. 
Since the dark sector is partially thermalized via gravitational interactions with the SM sector, 
%thermal-decoupled SM and dark sectors, 
%therefore the $t$ and $T$ are related by the Hubble parameter $H$ according to
%\begin{eqnarray}
%dt = -{\frac{dT}{T H(T)}}\,,~~~~~
%%t-t_c=-\frac{1}{H(T)}\ln \frac{T}{T_c}\,,~~~~
%H^2(T) =  \frac{\rho(T)}{3 M^2_{\rm Pl}}\,,
%\end{eqnarray}
the total radiation energy density is $\rho(T) = \frac{\pi^2}{30} g_\ast T_{\rm SM}^4$, where $g_\ast = g^{\rm SM}_\ast +g^{\rm D}_\ast\left(T/T_{\rm SM}\right)^4$ is the total number of relativistic degrees of freedom when the dark and SM sectors are at temperatures $T$ and $T_{\rm SM}$, 
respectively, and $g^{\rm D}_\ast =4.5$ at all relevant times.
We properly account for the amount of dark radiation after the FOPT by relating 
the effective number of extra neutrino species contributed by the dark sector after the phase transition, $\Delta N_{\rm eff}$, to $T/T_{\rm SM}$ at the end of the FOPT~\cite{Marfatia:2021twj}.

We identify the phase transition temperature $T_\star$ with the temperature of percolation, which we define as the temperature at which a fraction $1/e$ of the space is in the false vacuum.
Then, the time of the phase transition $t_\star$ is given by
\begin{eqnarray}
F(t_\star)=1/e\simeq 0.37 \,.
\label{cond}
\end{eqnarray}

As the FOPT proceeds, the false vacuum separates into smaller volumes. The formation of FBs
in a {\it false vacuum bubble} at $T_\star$ dominates over true vacuum bubble nucleation inside it
when the false vacuum bubble reaches a critical volume $V_\star$ given by $\Gamma(T_\star)V_\star R_\star \sim v_w$~\cite{Hong:2020est}. Here, $R_\star$ is the radius of the false vacuum bubble and $v_w$ is the bubble wall velocity.
Assuming that each critical volume results in one FB,
the number density of FBs is determined by
$n_{\rm FB}|_{T_\star} V_\star = F(t_\star)$.
%\begin{eqnarray}
%\label{eq:nFB}
%n_{\rm FB}|_{T_\star}= \left( \frac{3}{4\pi} \right)^{1/4}
%\left( \frac{\Gamma(T_\star)}{v_w} \right)^{3/4} F(t_\star)\,.
%\end{eqnarray}
%

For stable FBs to exist in the false vacuum, during the phase transition
there must be a nonzero asymmetry $\eta_\chi \equiv (n_\chi-n_{\bar{\chi}})/s$ 
in the number densities
(where the entropy density is  $s=(2\pi^2/45)\left(g^{\rm SM}_{*s}\,T_{\rm SM}^3+g^{\rm D}_{*s}\,T^3\right)$), so that a net number of particles remain after pair annihilation 
$\bar{\chi}\chi \to \phi \phi$, and 
$\chi$ must carry a conserved global $U(1)_Q$ so that 
$Q$-charge can be accumulated to stabilize a FB~\cite{Hong:2020est}.
The net $Q$-charge trapped in a FB, which is equivalent to the total number of $\chi$ in a FB, is 
%\begin{eqnarray}
$Q_{\rm FB}= \eta_\chi (  s/{n_{\rm FB}})_{T_\star}$.
%\end{eqnarray}
To be stable, the FBs must satisfy~\cite{Huang:2022him}
$$
\frac{dM_{\rm FB}}{dQ_{\rm FB}} < m+g_\chi v_\phi(T)\,,
~~{\rm and}~~\frac{d^2M_{\rm FB}}{dQ_{\rm FB}^2}< 0\,.
$$

Since a black hole violates continuous global symmetries, PBH formation via FB collapse will not conserve the $Q$-charge of the FB, and the PBH will evaporate as a Schwarzschild black hole.

Complete expressions for the energy, mass and radius of a FB are provided in appendix~\ref{sec:FB_profile}. Note that if the Yukawa energy is negligible, then for $0.1 \leq B^{1/4}/{\rm MeV} \leq 10^4$,  the FB radius is typically much larger than the Schwarzschild radius of a FB, and the FB does not collapse to a PBH.

%The total energy of a FB is given in Eq.~(\ref{eq:FB_energy}),
%\begin{eqnarray}
%\label{eq:FB_E}
%E= \frac{3\pi}{4} \left( \frac{3}{2\pi} \right)^{2/3} 
%\frac{Q^{4/3}_{\rm FB}}{R} + \frac{4\pi}{3}V_0(T_\star) R^3\,,
%\end{eqnarray}
%including the kinetic energy of Fermi-gas, Yukawa energy 
%and vacuum energy $V_0(T_\star) \equiv  V_{\rm eff}(0, T_\star)-V_{\rm eff}(v_\phi,  T_\star)$.
%
%The mass $M_{\rm FB}$ and radius $R_{\rm FB}$ of a FB are obtained by minimizing the FB energy with respect to its radius, and their analytical expressions are listed Eq.(\ref{eq:FB_mass_radius}).

%\begin{eqnarray}
%\label{eq:FB_mass_radius}
%M_{\rm FB}(T_\star)&=& Q_{\rm FB}(12\pi^2 V_0(T_\star))^{1/4} \,, \nonumber \\
%R_{\rm FB}(T_\star)&=& Q^{1/3}_{\rm FB}
%\left[ \frac{3}{16} \left( \frac{3}{2\pi} \right)^{2/3} \frac{1}{V_0(T_\star)} \right]^{1/4}\,.
%\end{eqnarray}
%
%If we neglect the Yukawa energy for the FB energy, the simplified formulas 
%for $M_{\rm FB}$ and $R_{\rm FB}$ are written in Eq.(\ref{eq:FB_mass_radius_simple}).
%For the energy range we are instrested in ($0.1 \leq B/{\rm MeV} \leq 10^4$),
%the In other words, without the Yukawa interaction, a FB hardly collapses into a PBH.

\subsection{FB collapse to PBH}

The $\chi$'s inside a FB  couple via the attractive Yukawa interaction 
$g_\chi \phi \bar{\chi}\chi$, 
with interaction length,
\begin{eqnarray}
L_\phi(T) \equiv 
\left( \left. \frac{d^2V_{\rm eff}}{d\phi^2}\right|_{\phi=0} \right)^{-1/2}
%= M^{-1}_\phi 
=\left(2D(T^2-T^2_0) \right)^{-1/2}\,.
\end{eqnarray}
The Yukawa potential energy contribution $E_Y$  to the total FB energy in Eq.~(\ref{eq:FB_energy}) is
\begin{eqnarray}
E_Y\simeq -\frac{3g^2_\chi}{8 \pi} \frac{Q^2_{\rm FB}}{R}\,  \left(\frac{L_\phi}{R} \right)^2\,.
\end{eqnarray}
As we can see, $L_\phi(T)$ and $|E_Y|$ increase
as the temperature decreases.
When $|E_Y|$ becomes larger than the Fermi-gas kinetic energy, the FB becomes unstable
and starts collapsing to a PBH. This coincides with $L_\phi$ becoming roughly the same as the mean separation distance of $\chi$'s, i.e., $L_\phi\simeq R_{\rm FB}/Q^{1/3}_{\rm FB}$.
More specifically, the temperature $T_{\phi}$ for PBH formation is
given by~\cite{Kawana:2021tde}
\begin{eqnarray}
\label{eq:1_3}
L_\phi(T_{\phi})=\left(2D(T^2_{\phi}-T^2_0) \right)^{-1/2}=
\frac{1}{g_\chi} \sqrt{\frac{2\pi}{3\sqrt{3}}}
\left( \frac{2\pi}{3} \right)^{1/6}
\frac{R_{\rm FB}}{Q^{1/3}_{\rm FB}}\,.
\end{eqnarray}
For our energy scale of interest ($0.1 \leq B^{1/4}/{\rm MeV} \leq 10^4$)
we find $T_\star > T_{\phi}$, so that the FB forms first and then collapses to a PBH. 
Immediately after formation, the PBH mass is obtained by
replacing $T$ with $T_{\phi}$ in Eq.~(\ref{eq:FB_mass_radius}),  
i.e., $M_{\rm PBH}=M_{\rm FB}(T_{\phi})$,
and the number density is $n_{\rm PBH}=n_{\rm FB}|_{T_\star}s(T_{\phi})/s(T_\star)$ 
due to the adiabatic evolution of the Universe~\cite{Kawana:2021tde}.
Note that since the FBs have a monochromatic mass function, so do the PBHs.
The PBH number density and relic abundance in the present Universe 
are given by~\cite{Carr:2020gox}
\begin{eqnarray}
\label{eq:DM_relic}
{n_{\rm PBH}|_0 \over s_0}&=& \left( \frac{n_{\rm PBH}}{s} \right)_{T_{\phi}}={3\over 4} {T_\phi \over M_{\rm PBH}}\frac{\rho_{\rm PBH}(T_{\phi})}{\rho(T_{\phi})}\,, \\
\Omega_{\rm PBH}h^2 &=& \frac{M_{\rm PBH}|_0\, n_{\rm PBH}|_0 }{3 M^2_{\rm Pl}(H_0/h)^2}\,,
\end{eqnarray}
where the Hubble constant, $H_0=2.13h\times 10^{-42}$~GeV.

\bigskip

\section{PBH evaporation spectra}
\label{sec:PBH_evaporation}

%If the PBHs constitute hundred percent of DM relic density,  we found the anticorrelation between energy scale of FOPT ($B^{1/4}$) and the $M_{\rm FB}$~\cite{Marfatia:2021twj}.
%Therefore for $0.1 \leq B^{1/4}/{\rm MeV} \leq 10^4$, the corresponding PBH mass range
%$10^{-22} \lesssim  M_{\rm PBH}/M_\odot \lesssim 10^{-9}$.
%

The evaporation of a PBH produces a primary component of particles from Hawking emission
and a secondary component from their decay and fragmentation. 
Hawking emission includes all particles with mass below the PBH temperature, 
regardless of their quantum numbers.
A PBH emits primary particles thermally with temperature $T_{\rm PBH}=M_{\rm {Pl}}^2/M_{\rm PBH}$, 
and numerically expressed as
\begin{eqnarray}
T_{\rm PBH}\simeq 5.3~{\rm MeV}  \times
\left( \frac{10^{-18} M_\odot}{M_{\rm PBH}} \right)\,.
\end{eqnarray}
For $M_{\rm PBH}/M_\odot \lesssim 10^{-20}$, PBHs have evaporated before today. 
The emission rate of primary particle $i$ is given by~\cite{Hawking:1974rv,Hawking:1975vcx}
\begin{eqnarray}
\frac{dN_i}{dE dt}= 
\frac{n^{\rm d.o.f}_i  \Gamma_i(E,M_{\rm PBH})}{2\pi (e^{E/T_{\rm PBH}}\pm 1)}\,,
\end{eqnarray}
where $n^{\rm d.o.f}_i$ is the number of degrees of freedom of particle $i$,
the graybody factor $\Gamma_i(E,M)$ describes a wave packet scattering 
in the nontrivial PBH spacetime geometry from the PBH horizon to an observer at infinity, 
and the $+(-)$ in the denominator corresponds to fermions (bosons). 

\begin{figure}[t!]
\centering
\includegraphics[height=2.8in,angle=270]{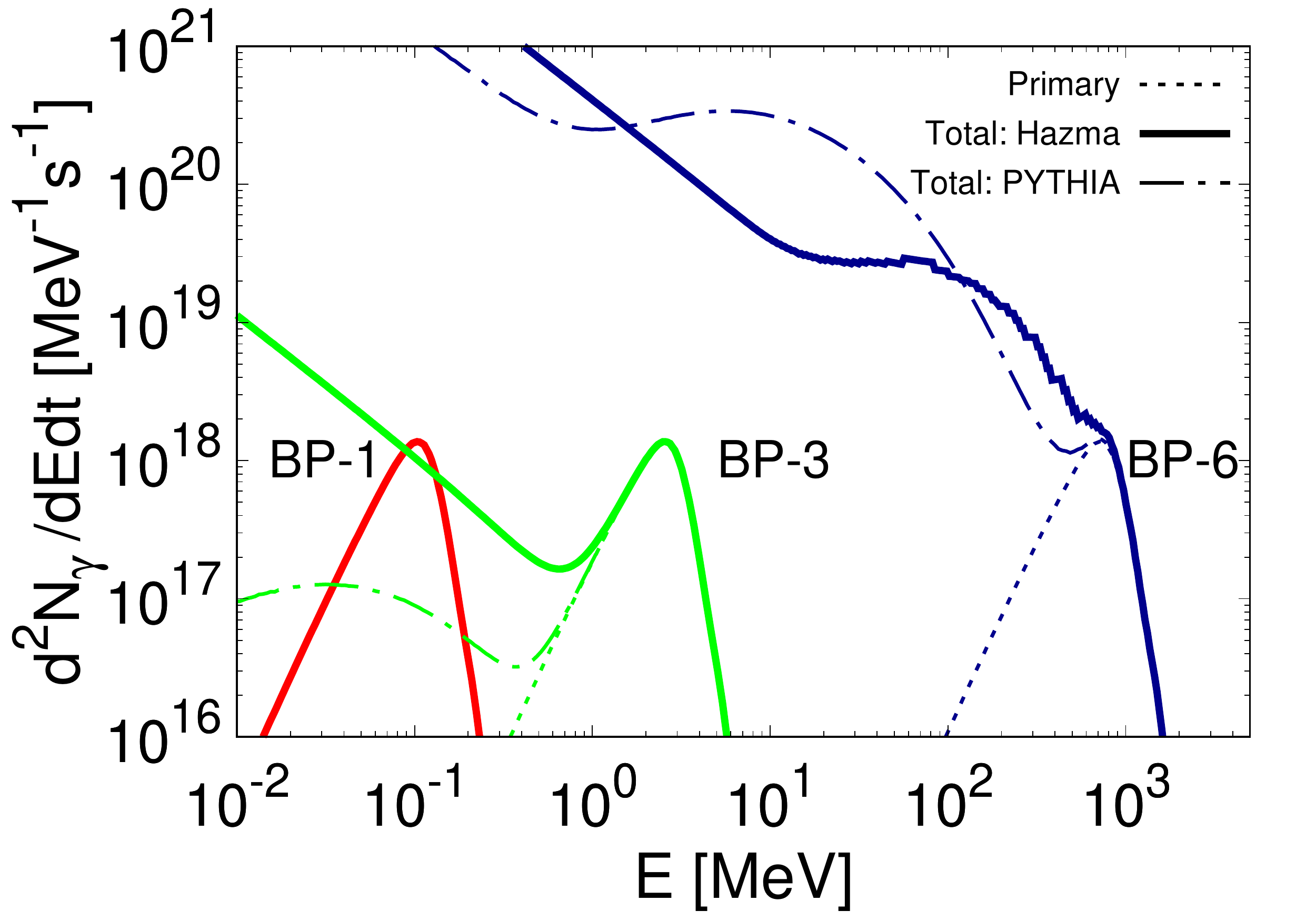}
\includegraphics[height=2.8in,angle=270]{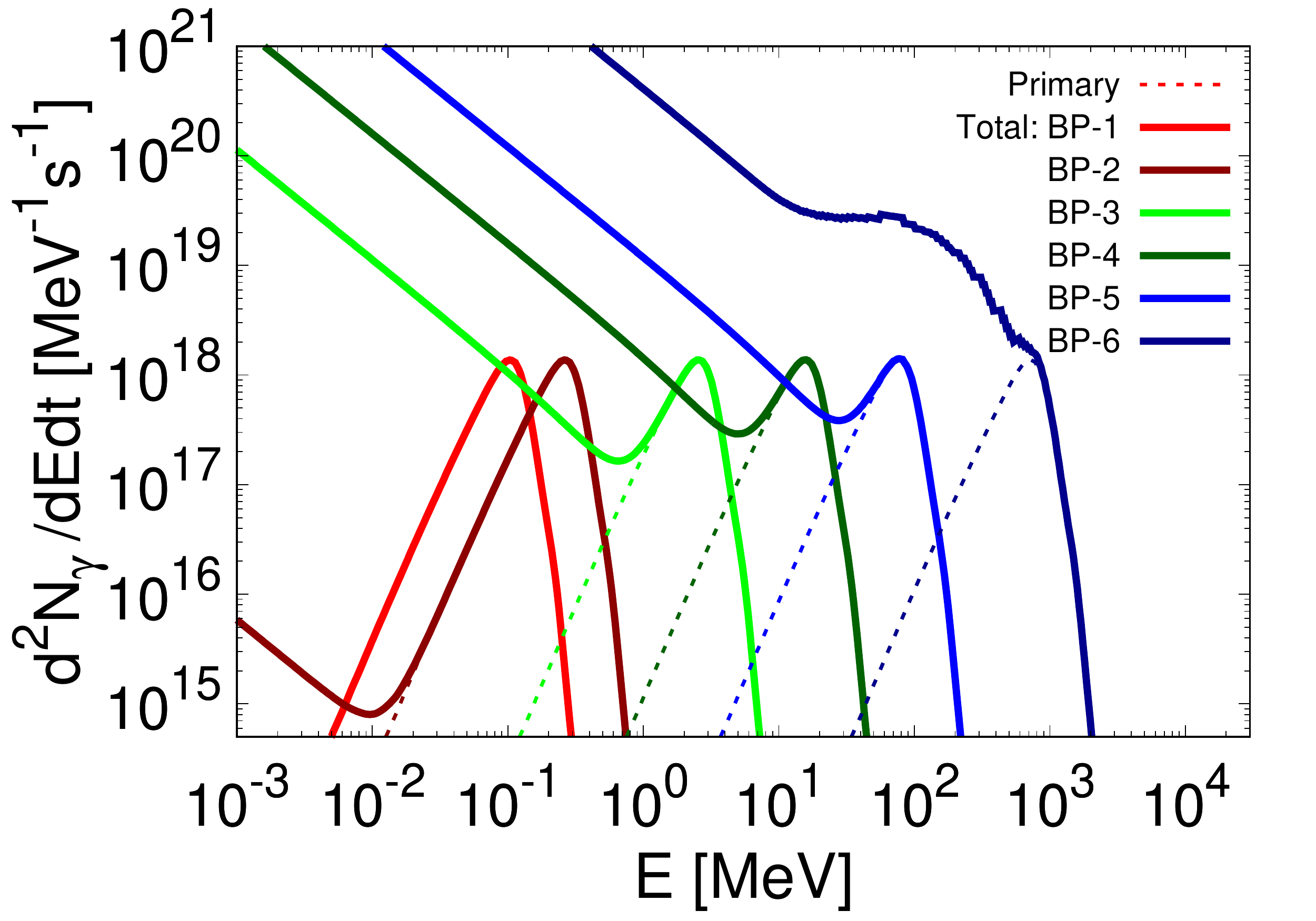}
\caption{\small \label{fig:evaporatin_PBH_tot}
Left panel: Instantaneous photon spectra from PBH evaporation using BlackHawk with Hazma and with PYTHIA for three benchmark points. The Hazma and PYTHIA spectra are identical for {\bf BP-1}, which has a relatively heavy PBH mass. Note the large differences in the secondary contributions for {\bf BP-3} and {\bf BP-6}.
Right panel: 
The photon spectra for all six benchmark points in Table~\ref{tab:BP}.
 %for PBH masses and $f_{\rm PBH}$ of {\bf BP-1} to {\bf BP-6}. 
}
\end{figure}

We restrict our study to photon emission 
because it provides the most severe model-independent constraints on our parameter space.
In order to compare with observations, the secondary component of emitted particles must be carefully calculated. Particle decays, 
quarks hadronization and fragmentation have significant effects
on the spectra at low energies.
In particular, secondary photons are produced in final state radiation from, and decays of,
 primary particles ($e^\pm\,,\mu^\pm\,,\pi^\pm\,,\pi^0$).
We employ the software package BlackHawk v2.0~\cite{Arbey:2019mbc,Arbey:2021mbl}
with Hazma~\cite{Hazma,Coogan:2020tuf} to compute the photon spectra from
PBH evaporation. In Fig.~\ref{fig:evaporatin_PBH_tot} we display a comparison between the instantaneous photon spectra using Hazma and PYTHIA~\cite{Pythia}/HERWIG~\cite{Herwig}. Clearly, the secondary components
of the spectra are very different for PBHs lighter than $\sim 10^{-17} M_\odot$. The primary reason for this is that the PYTHIA/HERWIG  hadronization
procedure is not reliable below 5~GeV.

%\begin{figure}[t!]
%\centering
%\includegraphics[height=4.2in,angle=270]{FIG/PBH_evaporation/inst/E_a_HAZMA.pdf}
%\includegraphics[height=2.8in,angle=270]{FIG/PBH_evaporation/inst/E_e.pdf}
%\includegraphics[height=2.8in,angle=270]{FIG/PBH_evaporation/inst/E_DM.pdf}
%\caption{\small \label{fig:evaporatin_PBH}
%}
%\end{figure}

The differential full-sky extragalactic $\gamma$-ray background (EGB) flux
due to PBH evaporation is given by~\cite{Carr:2020gox}
\begin{eqnarray}
{\frac{d\Phi}{dE}}= 
n_{\rm PBH}|_0\int^{{\rm min}(t_{\rm eva}\,,t_0)}_{t_{\rm CMB}} 
 c[1+z(t)] 
 %\frac{f_{\rm PBH} \rho_{\rm DM}}{M_{\rm PBH}}
\left. \frac{d^2 N_\gamma}{d \tilde{E} dt}\right|_{\tilde{E}=[1+z(t)]E}dt\,.
\end{eqnarray}
%where $\rho_{\rm DM}=1.27~{\rm GeV\, m^{-3}}$ is the average dark matter density of the Universe~\cite{pdg_constant} and $f_{\rm PBH}$ is the fraction of DM constituted by
If the PBHs have a lifetime shorter than the age of the Universe, $n_{\rm PBH}|_0$ is interpreted as the number density today had the PBHs not evaporated.
The lower limit of the integral is the time of last scattering, $t_{\rm CMB}=3.8\times 10^5~{\rm yr}$, after which photons propagate freely. The upper limit is the smaller of lifetime of the PBH ($t_{\rm eva}$) and the age of the Universe ($t_0=13.77\times 10^{9}~{\rm yr}$).
We approximate the evolution of the Universe as matter dominated until the current epoch, so that 
\begin{eqnarray}
%&& 1+z_{\rm RD}(t)=\left(\frac{\color{red}t_{\rm uni}}{t} \right)^{1/2}\,,
%~~~~{\rm for}~t< 5\times 10^4~{\rm yr,}~~\text{radiation domination} \,, \nonumber \\
&& 1+z(t)=\left(\frac {t_0}{t} \right)^{2/3}\,.
\end{eqnarray}
%The instantanteous photon spectra for the benchmark points are shown in Fig.~\ref{fig:evaporatin_PBH_tot}.
%

From Eq.~(\ref{eq:DM_relic}), it is convenient to define the commonly used parameter~\cite{Carr:2020gox}
\begin{eqnarray}
\beta' & \equiv & \gamma^{1/2} \left( \frac{g_{*}(T_\phi)}{106.75} \right)^{-1/4}  \left({h\over 0.67}\right)^{-2}
\frac{\rho_{\rm PBH}(T_{\phi})}{\rho(T_{\phi})}\,,
\label{beta'}
\end{eqnarray}
which is a measure of the fraction of the energy density of the Universe in PBHs at formation. We set $h=0.67$. The parameter $\gamma$ is defined by $M_{\rm PBH} \equiv\gamma M_H$,  where $M_H=c^3t/G$ is the Hubble horizon mass in the radiation dominated era. We treat the definition of 
$\gamma$ as a parametric relation with no connection to the dynamics of gravitational collapse. This is to enable us to use bounds presented in terms of $\beta'$. In our scenario, the prefactor in Eq.~(\ref{beta'}) is determined by the calculable quantities, $M_{\rm PBH}$ and $T_\phi$:
\begin{equation}
\gamma^{1/2}  \left( \frac{g_{*}(T_\phi)}{106.75} \right)^{-1/4} = 4.58\times 10^{-12}{T_\phi \over {\rm MeV}} \left({M_{\rm PBH}\over 10^{-18} M_\odot}\right)^{1/2}\,.
\end{equation}

%In our scenario, PBHs are produced in a FOPT, so $f_{\rm PBH}$ and $\beta'$ are related to parameters that describe the FOPT~\cite{Kawana:2021tde}:
%\begin{eqnarray}
%f_{\rm PBH} & \simeq & 0.8\times 10^{-9} \times v^{-3}_w 
%\left( \frac{g_*(T_\phi)}{10} \right)^{1/2}
%\left( \frac{T_\star}{{\rm MeV}} \right)^3
%\left( \frac{\beta/H_\star}{10^3} \right)^3
%\frac{M_{\rm PBH}}{10^{-18} M_\odot}\,,\nonumber \\
%
%\beta' &\simeq & 1.2\times 10^{-27} \times v^{-3}_w 
%\left( \frac{g_*(T_\phi)}{10} \right)^{1/2}
%\left( \frac{T_\star}{{\rm MeV}} \right)^3
%\left( \frac{\beta/H_\star}{10^3} \right)^3
%\left(\frac{M_{\rm PBH}}{10^{-18}M_\odot} \right)^{3/2}\,,
%\end{eqnarray}
%where $\beta$ is the inverse duration of the phase transition, 
%and $\beta/H_\star$ is obtained from the effective potential~\cite{Marfatia:2021twj}.
%

We consider two types of bounds on $\beta'$. The diffuse
extragalactic $\gamma$-ray background measured by HEAO-1, COMPTEL, EGRET and Fermi-LAT
places a strong, secure bound for $10^{-19} \lesssim M_{\rm PBH}/M_\odot \lesssim 10^{-15}$~\cite{Carr:2020gox}. The damping of small scale CMB anisotropies by PBH evaporation after recombination places the most stringent constraints on PBHs with mass $10^{-20} \lesssim M_{\rm PBH}/M_\odot \lesssim 10^{-19}$~\cite{Carr:2009jm}. The combined bound is shown in 
the $(M_{\rm PBH}/M_\odot, \beta')$ panel of Fig.~\ref{fig:GW_PBH} as a solid black curve.
%
%Futher more, the photons from PBHs evaporations, 
%with masses $10^{-22} \lesssim M_{\rm PBH}/M_\odot \lesssim 10^{-20}$, 
%can induce distortions in the CMB spectrum~\cite{Carr:2020gox}.
%
%Finally, the BBN puts bounds on $10^{-24} \lesssim M_{\rm PBH}/M_\odot \lesssim 10^{-20}$
%, because PBH evaporations during cosmological nucleosynthesis change the baryon-to-entropy ratio, resulting in increase of $^4$He but decrease the Deutron abundances~\cite{Miyama:1978mp}. 

\section{Correlated signals}
\label{sec:scan}

\begin{figure}[t!]
\centering
\includegraphics[height=1.5in,angle=0]{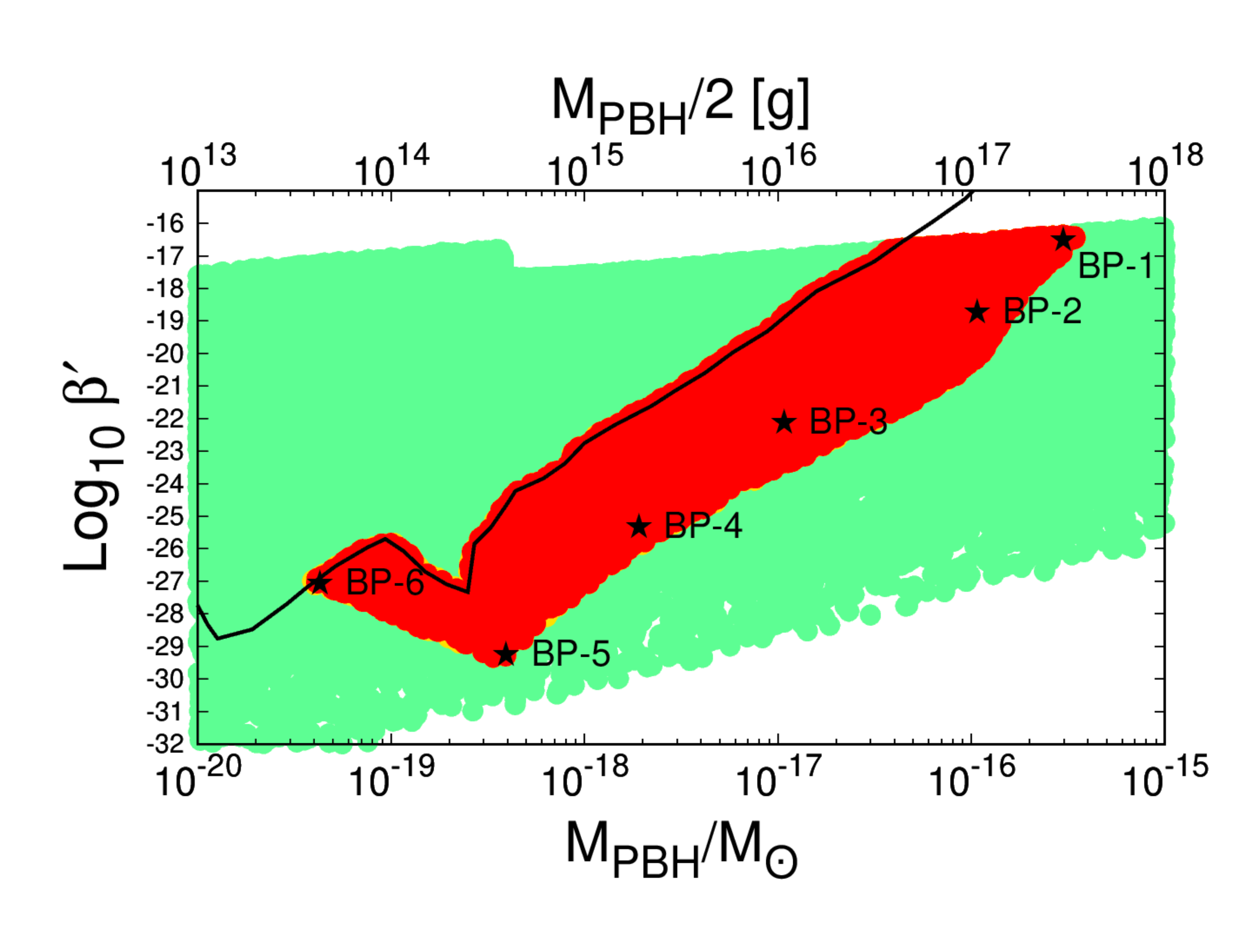}
\includegraphics[height=1.5in,angle=0]{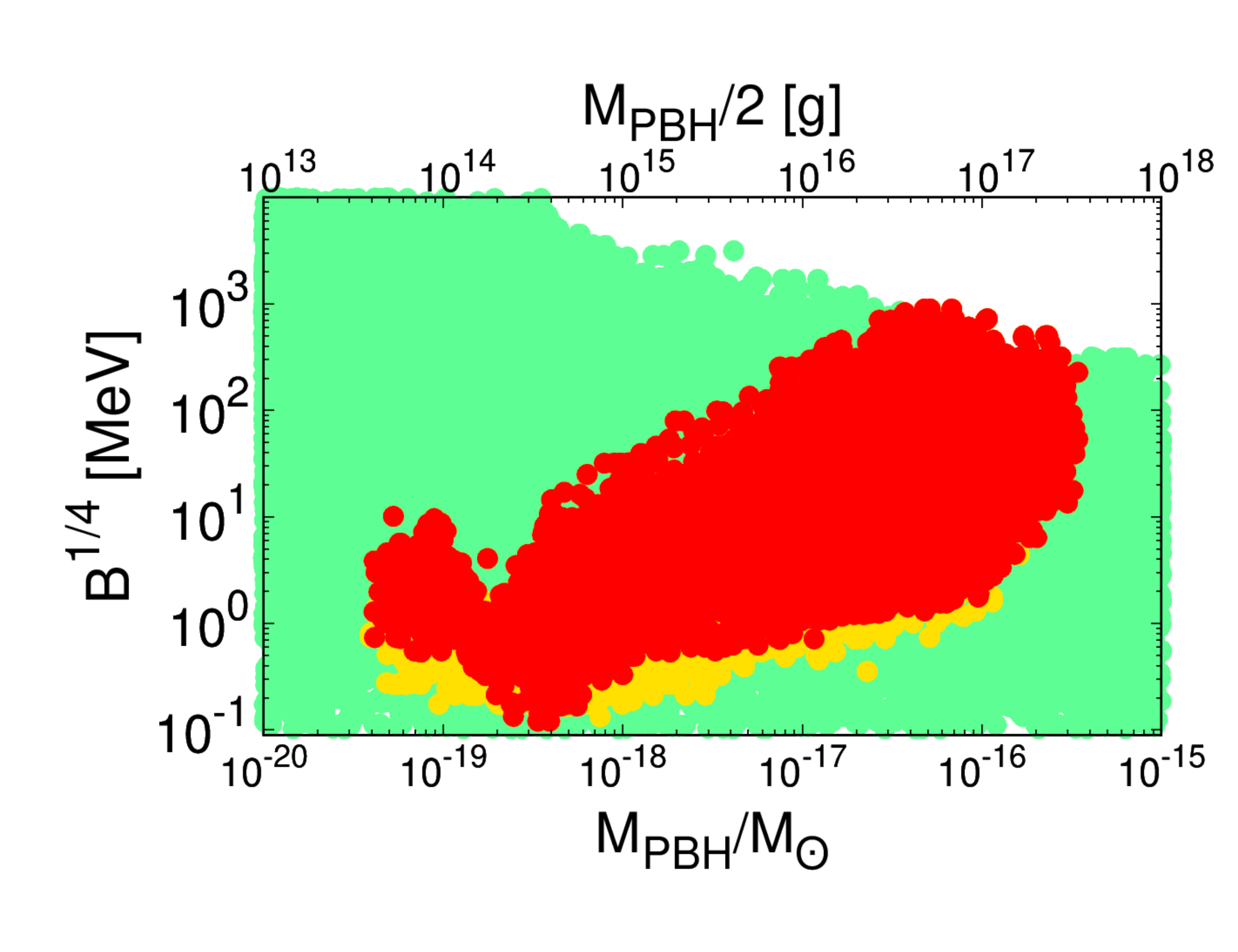}
\includegraphics[height=1.5in,angle=0]{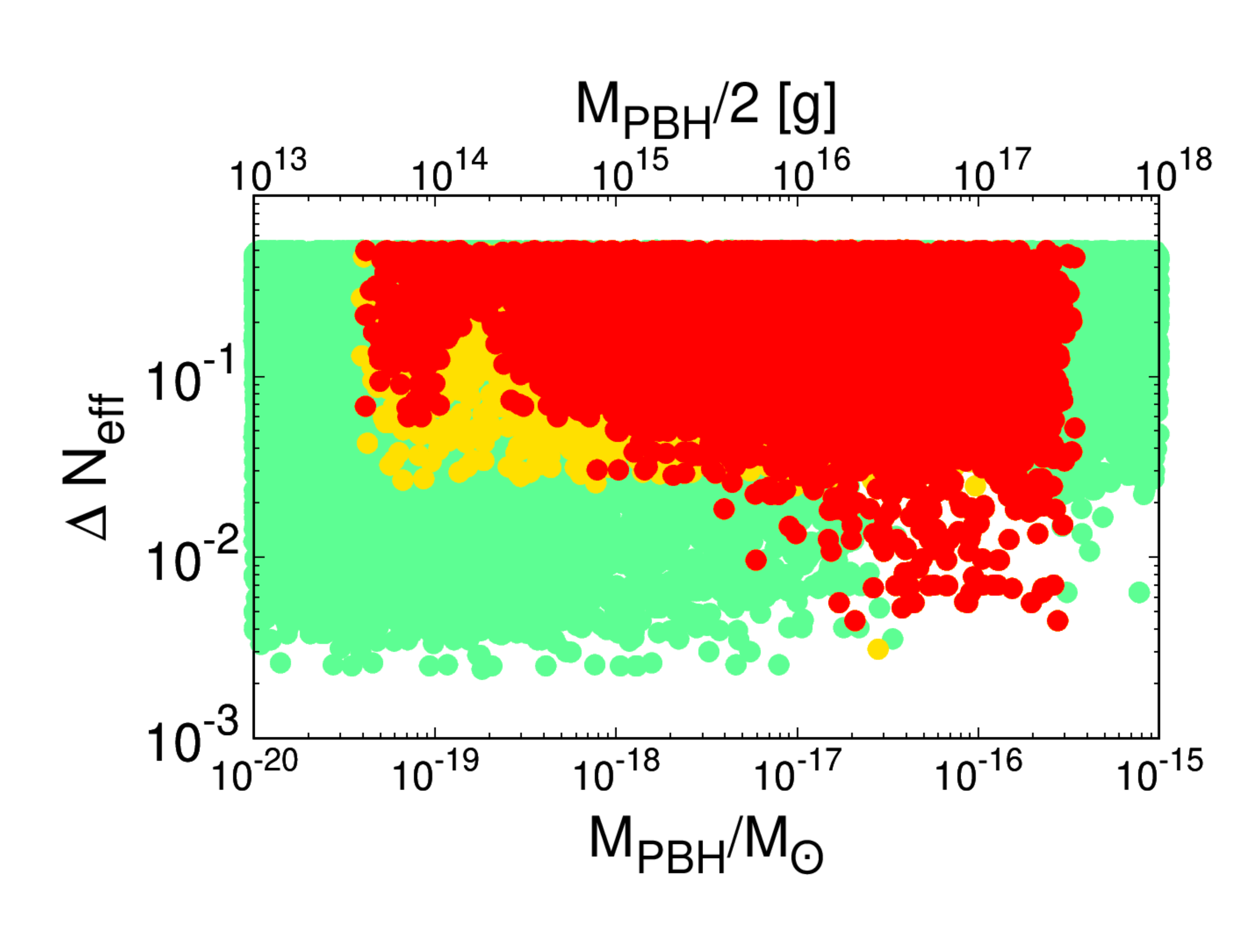}
\includegraphics[height=1.5in,angle=0]{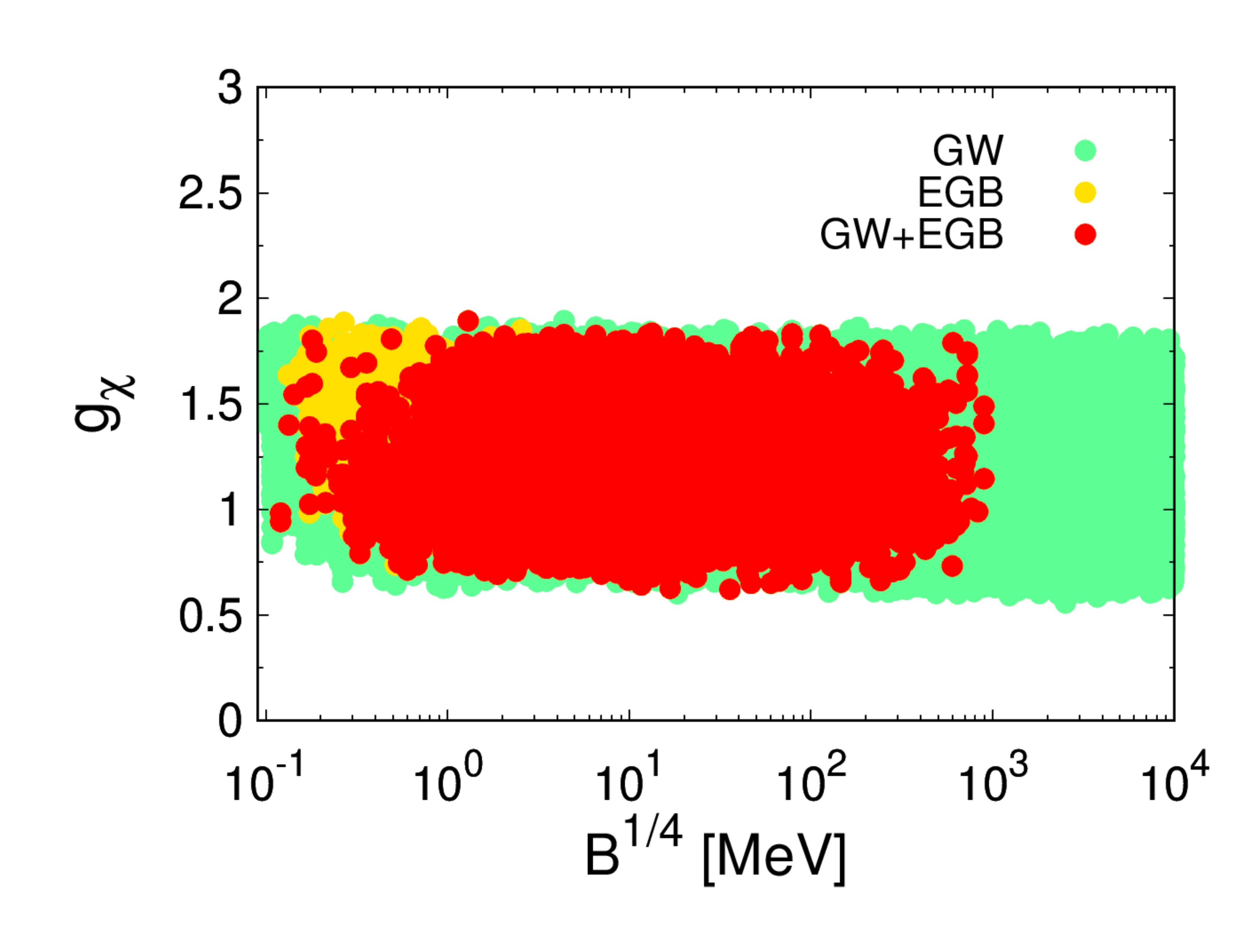}
\includegraphics[height=1.5in,angle=0]{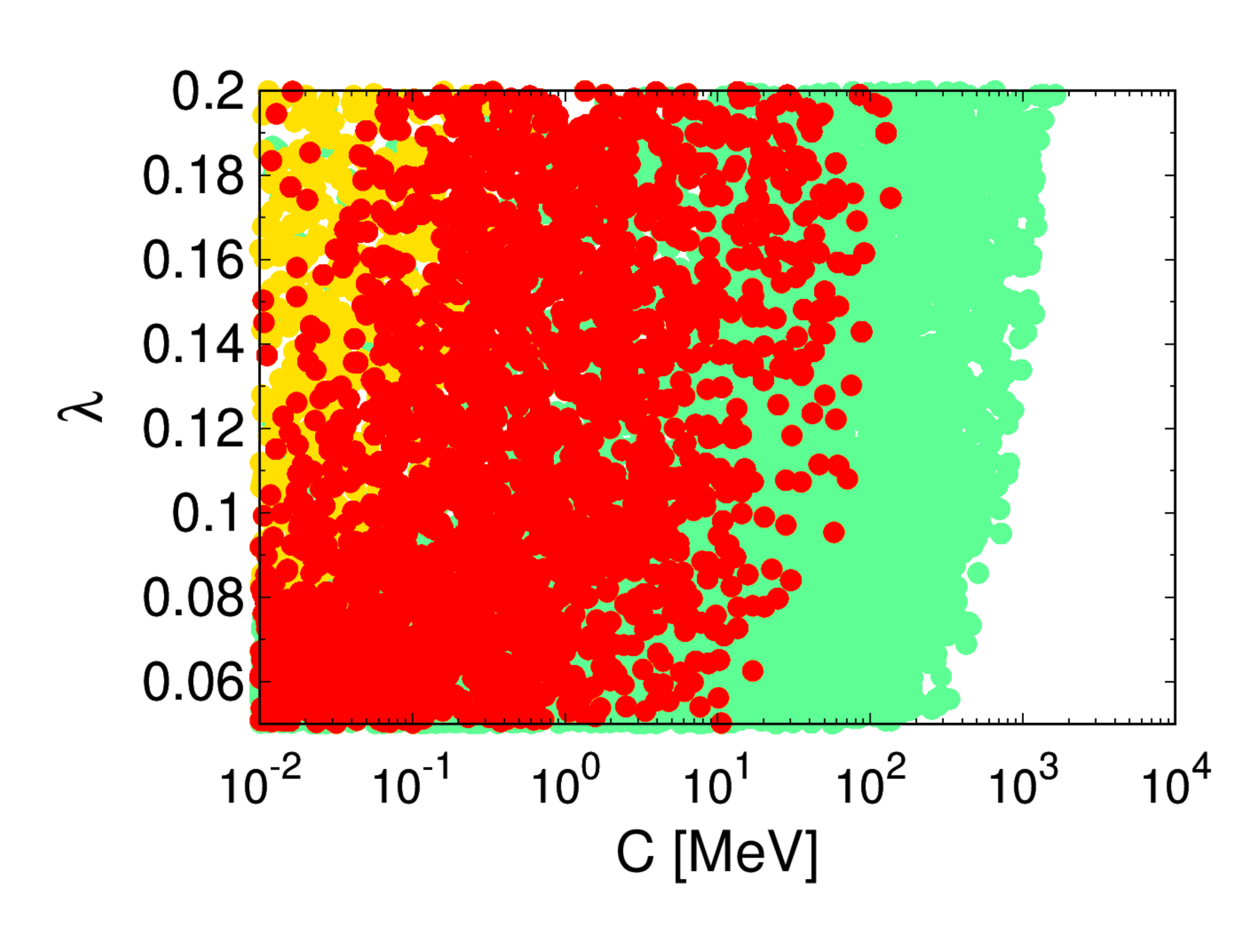}
\includegraphics[height=1.5in,angle=0]{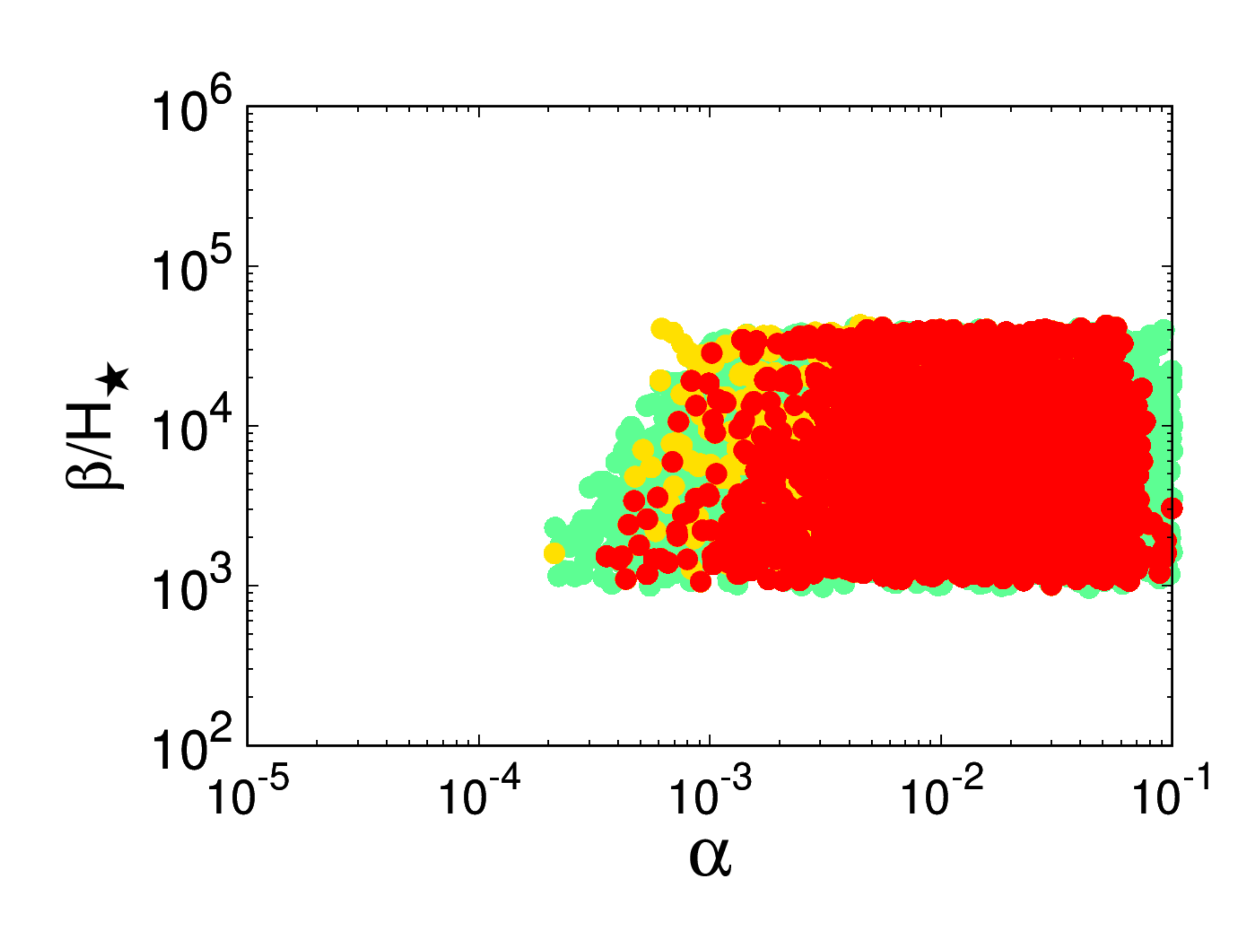}
\caption{\small \label{fig:GW_PBH}
The regions of parameter space that produce a detectable diffuse extragalactic X-ray/$\gamma$-ray background at  the MeV $\gamma$-ray telescopes, AMEGO-X/e-ASTROGAM (yellow), and a gravitational wave signal at THEIA/$\mu$Ares (green). 
In the red regions both measurements can be made.
The solid black curve in the top left panel is the current bound from observations of the extragalactic $\gamma$-ray background and from the damping of small scale CMB anisotropies.
All points satisfy $\Omega_{\rm PBH} h^2 \leq 0.12$ and $\Delta N_{\rm eff}\leq 0.5$.
}
\end{figure}

\begin{table}[t]
\caption{\small \label{tab:BP}
Benchmark points with $A=0.1$ fixed. $T_f$ is the temperature of the dark sector after the FOPT. 
$\alpha$ is the strength of the transition, defined by the ratio of the latent heat released and the
radiation energy density at the time of the FOPT, and $\beta$ is the inverse duration of the phase transition.}
%\begin{ruledtabular}
\begin{adjustbox}{width=\textwidth}
\begin{tabular}{c|cc|cc|cc}
\hline
\hline
    & {\bf BP-1} & {\bf BP-2} & {\bf BP-3} & {\bf BP-4} & {\bf BP-5} & {\bf BP-6}  \\
\hline
\hline
$\lambda$     
            & 0.061 & 0.110 & 0.195 & 0.087 & 0.150 & 0.158   \\
$B^{1/4}/{\rm MeV}$     
            & 75.14 & 13.81
            & 1.501 & 1.261
            & 0.121 & 2.999  \\
$C/{\rm MeV}$     
            & 0.249 & 0.462 & 0.078 & 0.052 & 0.011 & 0.325 \\
$D$         & 0.596 & 1.458 & 1.119 & 0.596 & 1.418 & 0.519  \\
$g_\chi$    & 1.088 & 1.301 & 1.011 & 1.289 & 0.983 & 1.228 \\
$\eta_\chi$     
            & $1.03\times 10^{-9}$ & $1.28\times 10^{-10}$
            & $1.64\times 10^{-12}$ & $1.21\times 10^{-15}$
            & $2.59\times 10^{-18}$ & $6.26\times 10^{-17}$  \\
$m/{\rm MeV}$   & 53.41  & 0.120 & 0.259 & 0.394 & 0.341 & 1.704 \\
%$r_T$       & 0.561 & 0.420 & 0.470 & 0.413 & 0.315 & 0.479 \\
\hline
$T_{{\rm SM}\star}/{\rm MeV}$     
            & 94.68 & 14.63 & 0.895 & 2.104 & 0.164 & 4.774 \\
$T_\star/{\rm MeV}$     
            & 53.16 & 6.143 & 0.421 & 0.868 & 0.052 & 2.287 \\
$T_f/{\rm MeV}$     
            & 59.63 & 6.888 & 0.472 & 1.023 & 0.068 & 2.571 \\
$T_{\phi}/{\rm MeV}$     
            & 53.09 & 6.045 & 0.415 & 0.857 & 0.050 & 1.950 \\
$S_3(T_\star)/T_\star$     
            & 155 & 159 & 166 & 171 & 180 & 170 \\
\hline
$M_{\rm PBH}/M_{\odot}$ 
            & $2.92\times 10^{-16}$ & $1.15\times 10^{-16}$
            & $1.19\times 10^{-17}$ & $1.93\times 10^{-18}$
            & $3.91\times 10^{-19}$ & $4.23\times 10^{-20}$ \\
$Q_{\rm FB}$   
            & $1.26\times 10^{42}$ & $4.31\times 10^{42}$
            & $5.96\times 10^{42}$ & $5.01\times 10^{41}$ 
            & $7.58\times 10^{41}$ & $4.18\times 10^{39}$ \\
$\beta'$ 
            & $2.80\times 10^{-17}$ & $2.54\times 10^{-19}$
            & $7.78\times 10^{-23}$ & $4.45\times 10^{-26}$
            & $5.75\times 10^{-30}$ & $8.97\times 10^{-28}$  \\
\hline
$\alpha$   
            & $1.48\times 10^{-2}$ & $7.40\times 10^{-3}$
            & $1.20\times 10^{-2}$ & $1.12\times 10^{-2}$ 
            & $1.35\times 10^{-2}$ & $1.30\times 10^{-2}$  \\
$\beta/H_\star$    
            & $4.41\times 10^{3}$ & $9.36\times 10^{3}$
            & $3.21\times 10^{4}$ & $3.25\times 10^{3}$ 
            & $4.94\times 10^{3}$ & $2.64\times 10^{3}$  \\
$v_w$                   
            & 0.904 & 0.904
            & 0.904 & 0.930
            & 0.963 & 0.905 \\
\hline
$v_\phi(T_\star)/{\rm MeV}$   & 224 & 23.1 & 1.426 & 3.821 & 0.247 & 8.157 \\
$dM_{\rm FB}/dQ_{\rm FB}/{\rm MeV}$   & 258 & 28.3 & 1.980 & 4.264 & 0.573 & 10.89 \\
\hline
$\Omega_{\rm PBH} h^2$
            & 0.079 & $1.12\times 10^{-3}$
            & $1.09\times 10^{-6}$ & $1.52\times 10^{-9}$ 
            & $2.15\times 10^{-13}$ & $6.35\times 10^{-29}$  \\
$\Delta N_{\rm eff}$
            & 0.218 & 0.126
            & 0.208 & 0.146
            & 0.147 & 0.221 \\
\hline
\hline
\end{tabular}
%\end{ruledtabular}
\end{adjustbox}
\label{2HDM}
\end{table}

We scan the parameters of the effective potential in Eq.~(\ref{eq:input}), 
the temperature ratio of the dark and SM sectors $T_\star/T_{\rm SM\star}$ during the phase transition,
the Yukawa coupling $g_\chi$, and the bare mass $m$ in the ranges,
\begin{eqnarray}
0.05 & \leq & \lambda \leq 0.2\,, \ \ \
0.1  \leq  B^{1/4}/{\rm MeV} \leq  10^4\,, \ \ \ 
0.01  \leq  C/{\rm MeV} \leq  10^4\,, \nonumber \\
0.1 & \leq & D \leq 10\,, \ \ \
0.3  \leq  T_\star/T_{\rm SM\star} \leq 1.0\,, \ \ \ \ \ \ \ \ \,
0.01  \leq  g_\chi \leq \sqrt{4 \pi}\,, \nonumber \\
10^{-3} & \leq &  m/B^{1/4} \leq  10\,.
\end{eqnarray}
We fix $A=0.1$ and adjust the value of the asymmetry parameter $\eta_\chi$
to ensure that $\Omega_{\rm PBH}h^2 \leq \Omega_{\rm DM}h^2\simeq 0.12$.
The results are shown in Fig.~\ref{fig:GW_PBH}.
In the green regions, GW signals from the FOPT are detectable
by the proposed THEIA~\cite{THEIA}, 
and $\mu$Ares~\cite{Sesana:2019vho} telescopes.
In the yellow regions, PBH evaporation produces an extragalactic X-ray/$\gamma$-ray
background that can be probed by 
future (AMEGO-X, e-ASTROGAM) 
and current (SPI, Fermi-LAT) $\gamma$-ray telescopes~\cite{e-ASTROGAM:2017pxr,Fleischhack:2021mhc}.
In the red regions, both the EGB and GWs can be observed. 

\begin{figure}[t!]
\centering
\includegraphics[height=4.2in,angle=270]{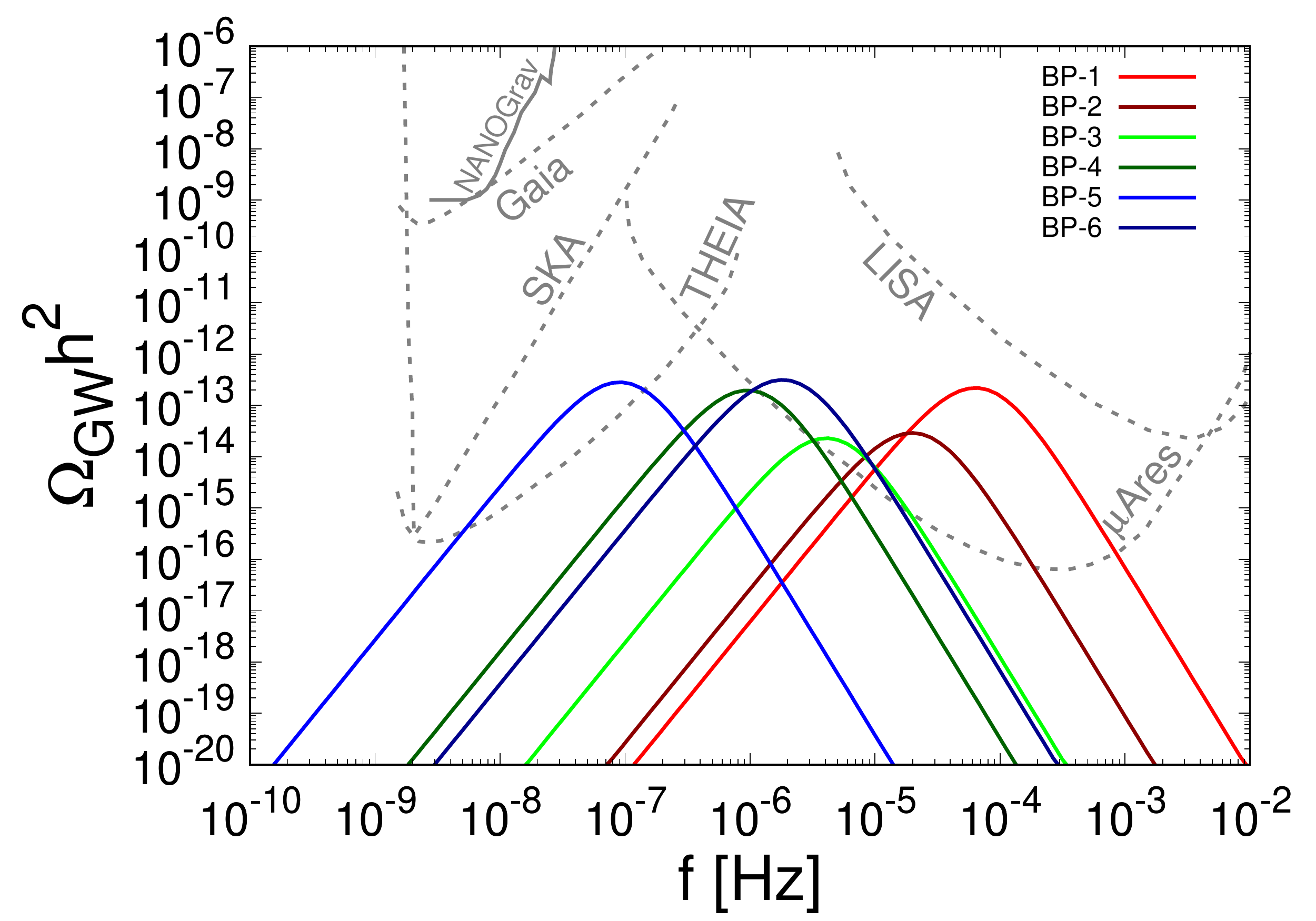}
\caption{\small \label{fig:GW}
Gravitational wave power spectra for the benchmark points in Table~\ref{tab:BP}.
}
\end{figure}

\begin{figure}[t!]
\centering
\includegraphics[height=4.2in,angle=270]{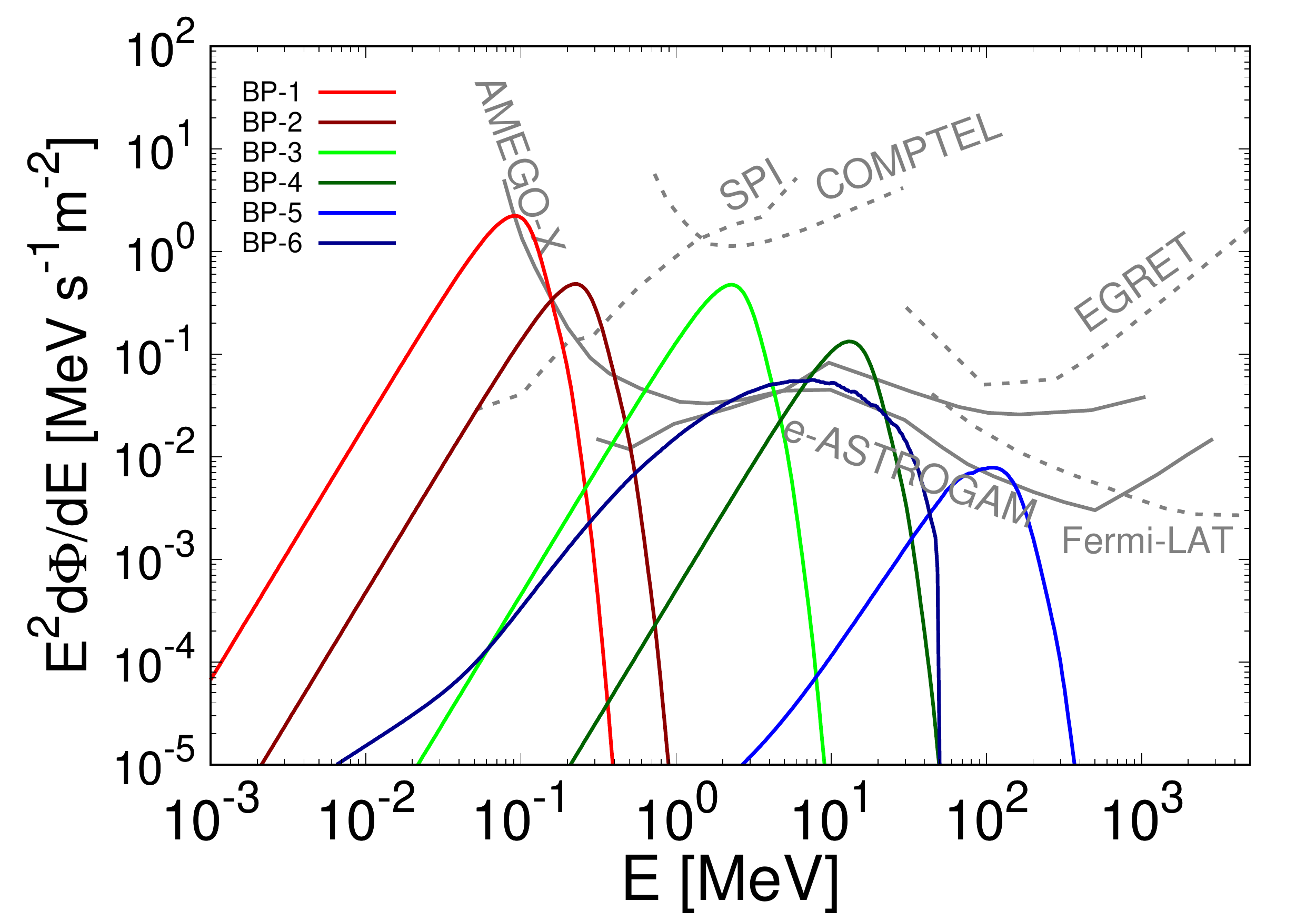}
\caption{\small \label{fig:gamma}
Extragalactic $\gamma$-ray spectra from PBH evaporation for the benchmark points in Table~\ref{tab:BP}.
}
\end{figure}

We select six benchmark points that lie in the red regions and that are compatible with the EGB and CMB bounds. They are  marked in the upper left panel of Fig.~\ref{fig:GW_PBH} and listed in Table~\ref{tab:BP}. %
Their GW spectra, arising from sound waves in the plasma during the FOPT, are shown in Fig.~\ref{fig:GW}; for the procedure used, see Ref.~\cite{Marfatia:2020bcs}.
$\mu$Ares is sensitive to all benchmark points except  {\bf BP-5}, which can be detected by
THEIA. We have also found points that can be detected by both $\mu$Ares and THEIA.
The corresponding EGBs are shown in Fig.~\ref{fig:gamma}. Our benchmark points are selected
to yield an observable EGB at the proposed AMEGO-X and e-ASTROGAM telescopes. The $3\sigma$ sensitivities, including those of the past and current telescopes, SPI, COMPTEL, EGRET, and Fermi-LAT are shown.
The spectral shape of {\bf BP-6} is different from that of the other points because in the instantaneous photon spectrum in Fig.~\ref{fig:evaporatin_PBH_tot},  the sharp peak expected from Hawking emission is washed out by secondary emission. Note that only {\bf BP-1} has  $\Omega_{\rm PBH}h^2 \sim\Omega_{\rm DM}h^2$, and that the relic density of {\bf BP-6} is vanishingly small because the PBHs have a much shorter lifetime ($\sim 10^{7}$~yr) than the age of the Universe.

\section{Summary}
\label{sec:summary}

We studied a novel mechanism of PBH formation, 
in which macroscopic Fermi balls produced during a FOPT in a dark sector become unstable and collapse to PBHs.
In the framework of a generic thermal effective potential, if the difference in vacuum energy at zero temperature is $0.1 \lesssim B^{1/4}/{\rm MeV} \lesssim 10^4$,
PBHs of mass $10^{-20}M_\odot \lesssim M_{\rm PBH} \lesssim 10^{-15}M_\odot$
are produced.

For $3\times 10^{-20}M_\odot \lesssim M_{\rm PBH} \lesssim 3\times 10^{-16}M_\odot$, the stochastic background of gravitational waves from the FOPT that peaks in the $10^{-7}$~Hz -- $10^{-4}$~Hz range can be detected at the proposed THEIA and $\mu$Ares telescopes, and in correlation,
a substantial extragalactic \mbox{X-ray/$\gamma$-ray} background produced by PBH evaporation can be detected at SPI, Fermi-LAT, and the proposed MeV $\gamma$-ray telescopes, AMEGO-X and e-ASTROGAM; see Figs.~\ref{fig:GW} and~\ref{fig:gamma}. A measurable amount of dark radiation is also typically expected; see Fig.~\ref{fig:GW_PBH}.

\bigskip
%\clearpage
\section*{Acknowledgements}  
We thank A.~Arbey and J.~Auffinger for help with the BlackHawk package. D.M. is supported in
part by the U.S. DOE under Grant No. de-sc0010504.

%\newpage
%%%%%%%%%%%%%%%%%%%%%%%%%%%%%%%
%%%%%%%%%%% Appendix %%%%%%%%%%%
%%%%%%%%%%%%%%%%%%%%%%%%%%%%%%%
\appendix
\section{FB properties}
\label{sec:FB_profile}

We derive analytic expressions for the FB mass and radius.
Including the Fermi gas kinetic energy, Yukawa potential energy, 
and the temperature-dependent potential energy difference between the false and true vacua,
the energy of a FB of radius $R$, temperature $T$, and charge $Q_{\rm FB}$, can be approximately written as~\cite{Kawana:2021tde}
\begin{eqnarray}
\label{eq:FB_energy}
E_{\rm FB}&=&
\frac{3\pi}{4} \left( \frac{3}{2\pi} \right)^{2/3} \frac{Q^{4/3}_{\rm FB}}{R}
\left[1+\frac{4 \pi}{9}\left(\frac{2\pi}{3} \right)^{1/3} \frac{R^2 T^2}{Q^{2/3}_{\rm FB}} 
\left(1+\frac{3}{2\pi^2}\frac{m^2}{T^2} \right)
 \right] \nonumber \\
&&
-\frac{3 g^2_\chi}{8 \pi} \frac{Q^2_{\rm FB} L^2_\phi}{R^3} 
+ \frac{4 \pi}{3} V_0 R^3 \left(1+\frac{T^2 m^2}{12 V_0} \right) \,, %\nonumber 
\end{eqnarray}
where $V_0(T)\equiv V_{\rm eff}(0,T) - V_{\rm eff}(v_\phi(T),T)$, which at zero temperature is $B$. Since the FB is a macroscopic object, a contribution from the surface tension is negligible.

To find $R_{\rm FB}$, we require $dE_{\rm FB}/dR=0$,
which yields the cubic equation,
\begin{eqnarray}
(R^2)^3+a_2(R^2)^2+a_1(R^2)+a_0=0\,,
\end{eqnarray}
where
\begin{eqnarray}
&& a_2 \equiv \frac{\pi}{12} \left( \frac{3}{2 \pi} \right)^{1/3}
\frac{Q^{2/3}_{\rm FB} T^2}{V_0}
\left(1+\frac{3}{2\pi^2}\frac{m^2}{T^2} \right)
\left(1+\frac{T^2 m^2}{12 V_0} \right)^{-1} \,, \nonumber \\
&& a_1 \equiv -\frac{3}{16} \left( \frac{3}{2\pi} \right)^{2/3}\frac{Q^{4/3}_{\rm FB}}{V_0}
\left(1+\frac{T^2 m^2}{12 V_0} \right)^{-1} \,, \nonumber \\
&& a_0 \equiv \frac{9 g^2_\chi}{32 \pi^2} \frac{Q^2_{\rm FB} L^2_\phi}{V_0}
\left(1+\frac{T^2 m^2}{12 V_0} \right)^{-1}
\,.
\end{eqnarray}
The largest of the three roots, 
$R^2_{\rm FB}=2\sqrt{q}\cos \theta-a_2/3$, 
gives the FB radius~\cite{Han:1999jc}. 
Then evaluating $E_{\rm FB}$ at $R_{\rm FB}$ gives the mass of the FB. We find
\begin{eqnarray}
\label{eq:FB_mass_radius}
R_{\rm FB} &=& Q^{1/3}_{\rm FB} \left[ \frac{1}{4}\left( \frac{3}{2\pi} \right)^{2/3} \frac{1}{V_0} \right]^{1/4} X^{1/2}\,,
\nonumber \\
M_{\rm FB} &=& \frac{3}{4} Q_{\rm FB} \left( 9 \pi^2 V_0 \right)^{1/4} X^{-3/2} \nonumber \\
&& \times \left\lbrace X+ \frac{4}{9}\left(1+\frac{T^2 m^2}{12V_0} \right)X^3 
+ \left(  \frac{2\pi}{9}\frac{T^2}{V^{1/2}_0}+\frac{1}{3\pi} \frac{m^2}{V^{1/2}_0} \right) X^2 
- \frac{2g^2_\chi L^2_\phi V^{1/2}_0}{3\pi} 
\right\rbrace\,, \nonumber \\
\end{eqnarray}
where $X$ is defined as~\cite{Han:1999jc}
\begin{eqnarray}
X&\equiv& \left[ \left( 1+ \frac{13}{108}\frac{T^2m^2}{V_0}+ \frac{\pi^2}{81} \frac{T^4}{V_0} + \frac{1}{36\pi^2}\frac{m^4}{V_0} \right)^{1/2} \cos\theta
-\frac{\pi}{18} \frac{T^2}{V^{1/2}_0}\left(1+\frac{3}{2\pi^2}\frac{m^2}{T^2} \right) \right] \left(1+\frac{T^2 m^2}{12V_0} \right)^{-1} \,,\nonumber \\ 
%&& {\rm and} \nonumber \\
\theta & \equiv & \frac{1}{3} \cos^{-1} \frac{r}{q^{3/2}}\,, \ \ {\rm with} \nonumber \\
r & \equiv & \frac{1}{6}(a_1a_2-3a_0)-\frac{1}{27}a^3_2 \nonumber \\
& = & -\frac{1}{256}\frac{Q^2_{\rm FB} T^2}{V^2_0}\left(1+\frac{T^2m^2}{12V_0} \right)^{-3} \nonumber \\
&& \times \left[ \left(1+\frac{3}{2\pi^2} \frac{m^2}{T^2}\right)
\left(1+\frac{35}{324}\frac{T^2m^2}{V_0}+\frac{2\pi^2}{243}\frac{T^4}{V_0}+\frac{1}{54\pi^2}\frac{m^4}{V_0} \right)
+ \frac{36 g^2_\chi}{\pi^2} \frac{ L^2_\phi V_0}{T^2}\left(1+\frac{T^2m^2}{12V_0} \right)^2
 \right]  \,, \nonumber \\
q & \equiv & -\frac{1}{3}a_1+\frac{1}{9}a^2_2
=\frac{1}{16} \left(\frac{3}{2\pi} \right)^{2/3}\frac{Q^{4/3}_{\rm FB}}{V_0}
\left(1+\frac{13}{108}\frac{T^2m^2}{V_0}+ \frac{\pi^2}{81} \frac{T^4}{V_0} + \frac{1}{36\pi^2}\frac{m^4}{V_0}  \right) \left(1+\frac{T^2m^2}{12V_0} \right)^{-2} \,. \nonumber \\
\end{eqnarray}
From the definition of $\theta$, a solution exists 
only if $|r/q^{3/2}| \le 1$.
If we neglect the Yukawa energy, i.e., set $a_0=0$, 
and Taylor expand $\cos \theta \simeq \frac{\sqrt{3}}{2}+\frac{r}{6q^{3/2}}$, we find
\begin{eqnarray}
 \cos\theta &\simeq & 
 \frac{\sqrt{3}}{2}-\frac{\pi}{36}\frac{T^2}{V^{1/2}_0}
\left(1+\frac{3}{2\pi^2}\frac{m^2}{T^2} \right)
\left(1+\frac{35}{324}\frac{T^2m^2}{V_0}+\frac{2\pi^2}{243}\frac{T^4}{V_0}+\frac{1}{54\pi^2}\frac{m^4}{V_0} \right) \nonumber \\
&&
\times \left(1+\frac{13}{108}\frac{T^2m^2}{V_0}+ \frac{\pi^2}{81} \frac{T^4}{V_0} + \frac{1}{36\pi^2}\frac{m^4}{V_0} \right)^{-3/2}\,, \nonumber \\
X &\simeq & \frac{\sqrt{3}}{2}\left(1+\frac{T^2 m^2}{12V_0} \right)^{-1}
\left(1+\frac{13}{108}\frac{T^2m^2}{V_0} + \frac{\pi^2}{81} \frac{T^4}{V_0} + \frac{1}{36\pi^2}\frac{m^4}{V_0} \right)^{-1} \nonumber \\
&& \times \left\lbrace \left(1+\frac{13}{108}\frac{T^2m^2}{V_0}
+ \frac{\pi^2}{81} \frac{T^4}{V_0} + \frac{1}{36\pi^2}\frac{m^4}{V_0} \right)^{3/2} \right. \nonumber \\
&& \left.-\left(\frac{\pi}{6\sqrt{3}} \frac{T^2}{V^{1/2}_0}+\frac{1}{4\sqrt{3}\pi}\frac{m^2}{V^{1/2}_0}\right)
\left(1+
\frac{113}{972}\frac{T^2m^2}{V_0}+\frac{8\pi^2}{729}\frac{T^4}{V_0}+\frac{2}{81\pi^2}\frac{m^4}{V_0} \right) 
 \right\rbrace \nonumber \\
&& \simeq 
\frac{\sqrt{3}}{2}\left(1-\frac{1}{4\sqrt{3}\pi}\frac{m^2}{V^{1/2}_0}
- \frac{\pi}{6\sqrt{3}}\frac{T^2}{V^{1/2}_0}-\frac{5}{216} \frac{m^2T^2}{V_0}  \right)\,.
\end{eqnarray}
In the limit, $V^{1/2}_0\gg T^2,m^2$, Eq.~(\ref{eq:FB_mass_radius}) can be simplified to
\begin{eqnarray}
\label{eq:FB_mass_radius_simple}
R_{\rm FB}&=& Q^{1/3}_{\rm FB} \left[ \frac{3}{16} \left( \frac{3}{2\pi} \right)^{2/3}
\frac{1}{V_0} \right]^{1/4}
\left(1-\frac{1}{8\sqrt{3}\pi}\frac{m^2}{V^{1/2}_0}
- \frac{\pi}{12\sqrt{3}}\frac{T^2}{V^{1/2}_0}-\frac{5}{432} \frac{m^2T^2}{V_0} \right)\,, \nonumber \\
M_{\rm FB} &=& Q_{\rm FB}\left( 12 \pi^2 V_0 \right)^{1/4}
\left(1+ \frac{\sqrt{3}}{8\pi} \frac{m^2}{V^{1/2}_0}
+\frac{\pi}{4\sqrt{3}}\frac{T^2}{V^{1/2}_0}-\frac{1}{16} \frac{m^2 T^2}{V_0} \right)\,.
\end{eqnarray} 

%%%%%%%%%%%%%%%%%%%%%-------------------

\end{document}